# Comment on "A glance beyond the quantum model"

Peter Morgan, Physics Department, Yale University, New Haven, CT 06520-8120.

**Summary. The aim of "A glance beyond the quantum model" to modernize the Correspondence Principle is compromised by an assumption that a classical model must start with the idea of particles, whereas in empirical terms particles are secondary to events. The discussion also proposes, contradictorily, that observers who wish to model the macroscopic world classically should do so in terms of classical fields, whereas, if we are to use fields, it would more appropriate to adopt the mathematics of random fields. Finally, the formalism used for discussion of Bell inequalities introduces two assumptions that are not necessary for a random field model, locality of initial conditions and non-contextuality, even though these assumptions are, in contrast, very natural for a classical particle model. Whether we discuss physics in terms of particles or in terms of events and (random) fields leads to differences that a glance would be well to notice.**

Navascués and Wunderlich (NW) present a modern formulation of the Correspondence Principle in the hope of being able to make general claims about the microscopic structure of the Universe, particularly with an eye on quantum gravity (Navascués and Wunderlich 2010). This is obviously a worthwhile enterprise in principle, but some of their assumptions about quantum theory, which are very widespread in the literature, are especially open to question when addressing such a general topic.

Although NW first introduces "particle talk" in the introduction, I begin with the following, "In a microscopic experiment of non-locality (see figure 1), there is an event in some intermediate region that produces a pair of particles", which is the second sentence of their Section 2. The second sentence of Section 3 adds that "a *macroscopic* experiment will start with a *macroscopic event* producing, not a pair of particles, but *N* independent pairs*,*" so there is no doubt that we are here talking about *particles*, whether we talk microscopically *or* macroscopically. It's a sad commentary on empiricist thinking that, when we talk about experiments, as we observe them macroscopically, with an eye on the Correspondence Principle, we don't begin with the observation that in modern quantum mechanical experiments we most typically observe *events*.

I will return below to how we can replace the particle talk with field talk. What particularly causes me to comment on NW is the assertion in Section 3 that "Now, in order to establish a connection with classical physics, Alice and Bob should not regard these intensities as fluxes of discrete particles, but rather as continuous fields." This is open to a severe technical criticism, that if Alice and Bob are any less classical than to think that the world is deterministic, they should definitely not think of the intensities as *continuous* fields, because continuous fields do not allow the mathematically rigorous addition of the concept of probability. If Alice and Bob make *any* observations that they report statistically, rather than as an unanalyzed list of times at which they observed millions of events, then in a foundational discussion in which they also introduce fields as a means to explain those statistics, they must consider how both fields and probability will be introduced.

Alice and Bob should instead regard the "intensities" as observables of a *random field*. There is a significant advantage in taking this approach, insofar as random fields admit presentation as operator-valued distributions, allowing a very close parallel with quantum fields. Indeed, the parallel is so close as to allow random fields to be constructed that are empirically equivalent to at least some quantum fields (Morgan 2009), although in a discussion that aims to present a modern correspondence principle a presentation of a random field as operator-valued distributions does not have to match a quantum field *that* closely. For discussions of random fields, see, for example, (Morgan 2009, 2007, 2006) or (Gudder 1979). A principle concern in using random fields is to present expectation values of various mutually commutative observables, which can be done very effectively by the construction of vector states in a Hilbert space.

Suppose now that we talk about events and fields instead of about particles. In classical terms, we observe "events", thermodynamic transitions of a small region of something like a photographic plate or a CCD, which are nonetheless still macroscopic or mesoscopic. Such events *would not occur if we did not put an essentially classical thermodynamically nontrivial object somewhere close to a state preparation apparatus*. From such an event, which is certainly not point-like when considered at a fine-grained level, we cannot warrant speaking of a particle as its cause. That even a closely correlated sequence of thermodynamic transitions in a cloud chamber cannot be taken as a warrant for talking about particles was shown very early in the history of quantum theory by Mott (1929), and we're all very used to loose particle talk causing trouble.

Consider a different classical sketch. We construct a thermodynamically metastable system that is so close to a transition point that it switches to a macroscopically different state (which we can either *see* happen or that can be recorded in a computer memory) even when it is in a shielded, darkened room. We implement a feedback mechanism, so that after this thermodynamic transition from the "ready" state to the "excited" state, we return the thermodynamic state to its ready state almost instantly. We call the number of transitions we see in the darkened room the "dark-rate" of the "detector". Even though the "detector" has detected nothing, it's necessary to have a non-zero dark-rate because there is a practical *and* principled trade-off of dark-rate with "sensitivity". When we move different preparation apparatuses into the room with the "detector", each of which we say generates a different field, the "detector" responds to each with a different rate of discrete thermodynamic transitions. Furthermore, if we have two or more detectors, we will observe different correlations and other higher-order statistics according to what field preparation apparatus we introduce.

These events and the events' statistics are not caused by any particles, so this sketch says, but because the detector is engineered to make thermodynamic transitions. Thermodynamic systems that are used as detectors are engineered to be maximally sensitive to changes in their environment. Hence, if the surrounding field that is their environment changes, the statistics change. Fortunately, the ways in which the statistics of thermodynamic transition events change as the field changes can be modeled successfully in a more-or-less instrumental way without modeling the detector as a statistical mechanical system.

Suppose now that we introduce other apparatus into the room, which variously reflects, refracts and otherwise transforms the field that is constructed by a preparation apparatus. We will observe that the rate of thermodynamic transitions of the detector will change, and we will observe various interference effects in the statistics as we move either the detector of any of the other apparatus, including changes of correlations and higher-order statistics. Given that it is the experimenter's job to create situations that are as weird as possible, hopefully weird enough to give theorists nightmares, how long can it be before the experimental statistical results are too weird to be described by a simple particle model? Even for a classical field, ray models cannot directly reproduce interference effects without significant additional mathematical structure.

There is, additionally, a statistical aspect to models of experiments — if we could construct effective classical continuous field models people would have been wise to this kind of thing long ago — the field must be subject to both quantum and thermal fluctuations, which requires the mathematics of random fields. The properties of the field have distinct scales that are very important. At a fine-grained scale, the field is definitely not at equilibrium, because it is in thermodynamic contact with one or several deliberately metastable detectors, but at a coarse-grained scale the field is definitely at equilibrium with the whole apparatus, in the sense that the statistics of the times at which thermodynamic transition events are observed must be time invariant (or, more generally, the experimenter must ensure that statistics are repeatable at different times). As a consequence, insofar as the system is in a coarse-grained equilibrium state, the statistics are conditioned by the whole experimental apparatus, in accordance with the strictures of the Copenhagen interpretation. I have presented evolving forms of this discussion in (Morgan 2009, 2007, 2006), so I will not further labor the account here. For other discussions of a field approach, see Hobson (2007) and Khrennikov (2009).

The discussion of macroscopic locality in Section 4 of NW should be compared with Bell's characterization of locality for beable theories (Bell 1976), as further discussed, in particular, by Shimony *et al.* (1993), d'Espagnat (1984), and Morgan (2006). I will first focus on NW's statement that "classical physics is a local theory" in the first paragraph of Section 4 as symptomatic of difficulties they have not faced, and that even Bell brushes off unwisely. We certainly expect the dynamics of a classical theory to be local, but *the initial conditions of a classical theory are unconstrained*, and are nonlocal by their very nature. Thus, on any given time-like hypersurface, we can freely introduce whatever nonlocal correlations are needed to describe what we observe at that time. Bell dismisses correlations in the initial conditions as a "conspiracy", and indeed a careful discussion is required (for which see Morgan 2006), but, for example, an equilibrium state at non-zero temperature already explicitly introduces nonlocal correlations. This alone is enough to problematize much of NW's discussion, but their construction of probabilities also takes instrument settings to be *parameters* instead of as *observables*. This change essentially requires that the description of measurement must be non-contextual instead of contextual, which is well-known to allow Bell inequalities to be derived without reference to locality (de Muynck 1986, Baez 1987, Landau 1987). Although NW's approach is a perfectly acceptable way to construct classical models when it is empirically adequate, when it fails we ought to include a model of the experimental apparatus. As discussed above, if we consider events to be caused by the joint action of a random field and thermodynamically metastable detectors, at coarse-grained equilibrium, then there is a clear possibility that the whole apparatus may have to be modeled to construct a good model, whereas if we consider events to be caused by two local particles emitted from a central source it seems absurd to consider the relevance of parts of an experimental apparatus that are far from a given particle. Contextual models are entirely natural for random field models, in contrast to being unnatural for classical particle models.

Navascués and Wunderlich have done something rather remarkable. By introducing the idea of continuous fields in their paper they have laid themselves open to a criticism that they must introduce random fields, and encourage a discussion that would otherwise have been impossible. If they had introduced fields without daring to go "beyond" the standard model, they would equally have been conventionally impervious. Navascués' and Wunderlich's paper requires a vigorous condemnation where something less ambitious would have gone unchallenged. Finally, this comment does not touch Navascués' and Wunderlich's argument; their paper's flaw, I think, and such little as it is, is to have introduced a classical field metaphysics and not to have thought enough of it. A flaw I know well.

# References

Baez, J. 1987 Bell's Inequality for C$^*$-Algebras. *Lett. Math. Phys.* **13**, 135-136. (doi: 10.1007/BF00955201)

Bell, J. S. 1987 *Speakable and Unspeakable in Quantum Mechanics*, Chapters 7 and 12. Cambridge: Cambridge University Press.

d'Espagnat, B. 1984 Nonseparability and the tentative descriptions of reality. *Phys. Rep.* **110** 201-264. (doi:10.1016/0370-1573(84)90001-2).

Gudder S., 1979, *Stochastic Methods in Quantum Mechanics*, Chapter 6. New York: North-Holland.

Hobson, A. 2007 Teaching Quantum Physics Without Paradoxes. *Physics Teacher* **45**, 96-99. (doi: 10.1119/1.2432086).

Khrennikov, A., 2009 Detection Model Based on Representation of Quantum Particles by Classical Random Fields: Born's Rule and Beyond. *Found. Phys.* **39**, 997-1022. (doi: 10.1007/s10701-009-9312-y).

Landau, L. J. 1987 On the violation of Bell's inequality in quantum theory. *Phys. Lett. A* **120**, 54-56. (doi:10.1016/0375-9601(87)90075-2).

Morgan, P. 2006 Bell inequalities for random fields. *J. Phys.* **A**: *Math. Gen.* **39**, 7441–7455. (doi:10.1088/0305-4470/39/23/018).

Morgan, P. 2007 Lie fields revisited. *J. Math. Phys.* **48**, 122302. (doi: 10.1063/1.2825148).

Morgan, P. 2009 Equivalence of the Klein-Gordon random field and the complex Klein-Gordon quantum field. *EPL* **87**, 31002. (doi: 10.1209/0295-5075/87/31002).

Mott, N. F. 1929 The wave mechanics of α-ray tracks. *Proc. R. Soc. Lond. A*, **126**, 79-84.

de Muynck, W. M. 1986 The Bell inequalities and their irrelevance to the problem of locality in quantum mechanics. *Phys. Lett. A* **114**, 65-67. (doi:10.1016/0375-9601(86)90480-9).

Navascués, M. & Wunderlich, H. 2010 A glance beyond the quantum model. *Proc. R. Soc. A* **466**, 881-890. (doi: 10.1098/rspa.2009.0453).

Shimony, A., Horne, M. A., and Clauser, J. F. 1993, *Search for a Naturalistic World View* Vol. 2 ed. A Shimony (Cambridge University Press, UK), Chapter 12.